\documentclass[]{krakproc}
\usepackage{psfig,wrapfig}
\begin{document}

\newcommand\HI{H\,{\sc i}}
\newcommand\HII{H\,{\sc ii}}
\def\etal{{\rm et~al.\ }}

\title{The Magnetic Field of the Large Magellanic Cloud: A~New Way of
Studying Galactic Magnetism}

\author{Bryan M. Gaensler\footnote{Harvard-Smithsonian Center
for Astrophysics, 60 Garden Street MS-6, Cambridge, MA 02138,
USA}~\footnote{School
of Physics, University of Sydney, NSW 2006, Australia}, M.
Haverkorn$^*$,
L. Staveley-Smith\footnote{Australia Telescope National
Facility, CSIRO, PO Box 76, Epping, NSW 1710, Australia}, J. M.
Dickey\footnote{Physics Department, University of Tasmania, GPO Box
252-21, Hobart, Tasmania 7001, Australia}, \\
N. M. McClure-Griffiths${^\ddagger}$,
J. R.  Dickel\footnote{Astronomy Department, University of Illinois, 1002
West Green Street, Urbana, IL 61801, USA}
~and M. Wolleben\footnote{Max-Planck-Institut f\"{u}r
Radioastronomie, Auf dem
H\"ugel 69, D-53121 Bonn, Germany}}
\institute{}

\maketitle

\vspace{-8mm}
\begin{abstract}
We present a study of the Faraday rotation of extragalactic sources
lying behind the Large Magellanic Cloud.  These data represent the most
detailed study yet of magnetic field structures in any external galaxy,
and are a demonstration of how magnetic fields in galaxies will best be
studied with future instruments.
\end{abstract}

\section{Introduction}

The Large Magellanic Cloud (LMC) 
is the prototype of the barred Magellanic
spiral class of galaxies, and shows spectacular evidence for on-going
interaction with both the Milky Way and the Small Magellanic Cloud
(SMC). The LMC's proximity, low inclination, and minimal foreground and
internal extinction all make it an ideal target for studying both specific
populations and overall galactic structure. There have correspondingly
been many recent detailed studies of stars, gas and dust in the LMC, but
surprisingly little is known about this galaxy's magnetic fields.  Here we
summarise previous work on the LMC's magnetism, present a new study of
the LMC using background rotation measures (RMs)
(see \cite{ghs+05} for further discussion), and explain how this
technique can be applied to this and other galaxies in future studies
(for a review see Gaensler \etal\ 2004\nocite{gbf04}
and \cite{bg04}).

\section{Previous Studies of the LMC's Magnetism}

The earliest studies of magnetic fields in the LMC were carried
out through the study of polarisation of optical starlight
(e.g., \cite{vis66}; \cite{mf70b}; \cite{sch76b}).  
These data suggested the presence of a possible
spiral field geometry in the vicinity of the star-forming region
30~Doradus, plus the overall presence of a ``pan-Magellanic field'',
directed along a line joining the LMC to the SMC (see \cite{way90}
for a summary).

The strength of the LMC's magnetic field can be estimated through
observations of its diffuse radio synchrotron emission: standard
equipartition arguments imply a mean field strength of $\approx6$~$\mu$G
(\cite{kwhm89}), a value supported by $\gamma$-ray observations which
allow a direct separation of energy in particles and in magnetic fields
(\cite{poh93}, but see also \cite{cw93}).  Additionally, one LMC pulsar
(PSR~B0529--66) has a measured RM, implying a line-of-sight field strength
in the range 0.5--2~$\mu$G (Costa \etal\ 1991\nocite{cmh91}).

\begin{figure}[thb]
\vspace{2mm}
\centerline{\psfig{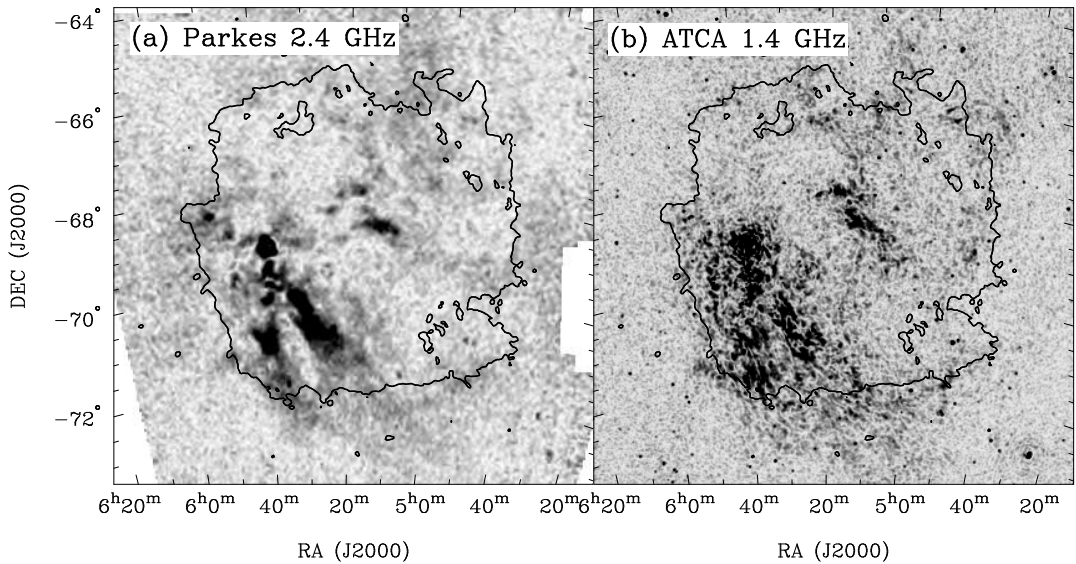}}
\caption{Linearly polarised radio emission from the LMC: 
(a) 2.4 GHz emission mapped by Parkes, smoothed to a resolution
of $30'$ (\cite{khwm93}); (b) 1.4 GHz emission mapped by the ATCA, smoothed
to a resolution of $3'$. In both cases, the contour
represents {\em IRAS}\ 60-$\mu$m emission at the level of
1~MJy~sr$^{-1}$, delineating the outer boundary of the LMC.}
\label{fig_diffuse}
\end{figure}

Direct imaging of the polarised synchrotron emission from the LMC
provides further clues as to this galaxy's magnetic field geometry.
Such imaging was carried out with the Parkes radio telescope by Klein
\etal\ (1993\nocite{khwm93}), revealing a polarised morphology which
was dominated by two highly polarised ``fingers'' of emission to the
south of 30~Dor, as shown in Figure~\ref{fig_diffuse}(a).  Klein \etal\
(1993\nocite{khwm93}) suggested that these might trace a magnetic loop,
extending out of the LMC's disk.

The vectors of polarised emission seen in this study suggested possible
spiral structure to the overall magnetic field, but uncorrelated with
the spiral arms seen in H$\alpha$ and in the mid-infrared. There
was no evidence for the pan-Magellanic field suggested by optical
data.  An important caveat to these results is that Klein \etal\
(1993\nocite{khwm93}) assumed that Faraday rotation toward the LMC was
small, and so did not correct for such effects.

\section{New Observations}

The LMC was surveyed in \HI\ by Kim \etal\ (1998,
2003\nocite{ksd+98,ksd+03}), using the Australia Telescope Compact Array
(ATCA). These observations involved approximately 1300 pointings at a
wavelength of 21~cm, the resulting image being sensitive to all spatial
scales in the range $1'$ to $30'$. The advantage of the ATCA is that
for this and many other \HI\ experiments, the correlator simultaneously
produced full multi-channel (14 $\times$~8-MHz) continuum polarimetry in
an adjacent frequency band. We have extracted these data from the archive
and have analysed the continuum component.  The resulting polarisation
data form an incredible resource for studying the LMC's magnetism at
unprecedented sensitivity ($\approx0.2$~mJy~beam$^{-1}$) and resolution
($\approx40''$).

\section{Results}

In Figure~\ref{fig_diffuse}(b) we show the resulting image of diffuse
emission (note that this image comes from an interferometer, and so
is not sensitive to structure on scales larger than $\sim30'$). The
two fingers seen with Parkes by Klein \etal\ (1993\nocite{khwm93}) are
clearly apparent in our new data. The rest of the polarised emission
from the LMC is comprised of possible filaments and shells of emission,
showing little correspondence with radio continuum, H$\alpha$, \HI\ or
mid-infrared emission.  These data will be fully discussed elsewhere;
clearly they indicate a complicated mixture of polarised emission and
depolarisation, as we have previously seen in our on-going studies of
our own Milky Way (e.g., \cite{gdm+00}; \cite{hkd03}).

\begin{figure}
\centerline{\psfig{file=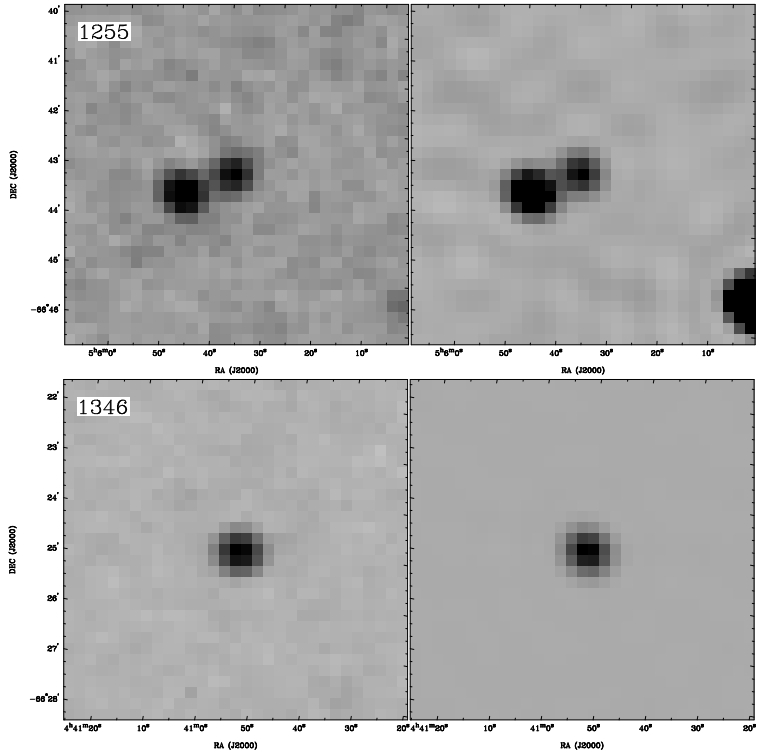,width=0.7\textwidth}}
\caption{Examples of linear polarisation from background sources in our
LMC survey.  In each case, the left panel shows the linearly polarised
intensity, while the right panel shows the total intensity (the intensity
ranges are on different scales for each panel).}
\label{fig_srcs}
\end{figure}

The image shown in Figure~\ref{fig_diffuse}(b) also demonstrates the
existence of linearly polarised emission from approximately 380 unresolved
background sources, examples of which are shown in Figure~\ref{fig_srcs}.
The typical polarised fraction from such sources is $\sim3$\%,
although the faintest sources have higher polarised fractions than
this due to sensitivity bias. While the total intensity image can be a
complicated mix of diffuse emission, SNRs and \HII\ regions, confusion
from diffuse emission is negligible in the polarised image, so that
even faint polarised sources can be easily identified (the image in
Fig.~\ref{fig_diffuse}[b] has been significantly smoothed).

\begin{wrapfigure}{r}{0.55\textwidth}
%\vspace{-2mm}
\centerline{\psfig{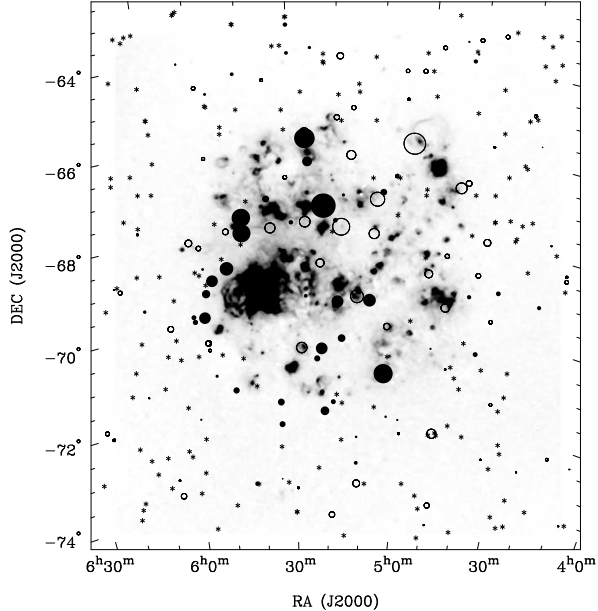}}
\caption{Distribution of RMs behind the LMC (\cite{ghs+05}). 
The greyscale denotes
emission measures derived from the SHASSA H$\alpha$ survey (\cite{gmrv01}),
while the closed and open circles represent positive and negative RMs,
respectively. Asterisks indicate RMs that are consistent with zero within
their errors.}
\label{fig_rms}
\vspace{-2mm}
\end{wrapfigure}

For each source, we obtain a measurement of the Stokes parameters $Q$
and $U$ in 14 independent channels centred on 1384~MHz.  Because these
channels are closely spaced, we do not suffer from the usual $n\pi$
ambiguities present in Faraday rotation measurements provided that RMs
have magnitudes less than $\sim2700$~rad~m$^{-2}$ (see Gaensler \etal\
1998\nocite{gmg98}).  For each frequency channel on each source, we
consider polarisation position angles for nine adjacent pixels. We fit
across the observing band for the RM, and then apply criteria for extent
of the source, signal-to-noise, quality of the fit, and pixel-to-pixel
scatter in the RMs (e.g., \cite{btj03}).  After all these tests were
applied, we obtained 292 accurate and reliable RMs in the field of the
LMC, at a source density of $\sim2.2$ measurements per square degree.

The resulting distribution of RMs is shown in Figure~\ref{fig_rms}, from
which a mean baseline RM has been subtracted from the data.  There is
a clear excess of Faraday rotation toward the LMC, demonstrating the
presence of a significant line-of-sight component to the LMC magnetic
field. The baseline-subtracted RMs through the LMC have a mean of
$+10$~rad~m$^{-2}$, but with a large scatter, values ranging from $-215$
to $+247$~rad~m$^{-2}$.  These RMs imply large rotations of polarisation
position angle in the data of Klein \etal\ (1993\nocite{khwm93}),
casting doubt on the validity of the magnetic field geometry that those
authors inferred.

\section{Discussion}

\subsection{The Coherent Magnetic Field}

We assume that the LMC disk is inclined to the plane of the sky by
35$^\circ$, with its line of nodes oriented at a position angle of
120$^\circ$, north through east (\cite{vdm04}). With this orientation,
the LMC RMs shown in Figure~\ref{fig_rms} can be shown to trace a
sinusoidal pattern as a function of position angle within the LMC's
disk, implying a vertically symmetric spiral magnetic field of mode
$m=0$. The pitch angle is difficult to constrain, but is likely to
be reasonably low, $\la20^\circ$.  The amplitude of our sinusoidal
fit is 53~rad~m$^{-2}$, while the equivalent dispersion measure,
as inferred from observations of LMC radio pulsars (\cite{ckm+01}),
is $\approx100$~pc~cm$^{-3}$. The implied strength of the uniform
field component in the LMC is $\approx1.1$~$\mu$G. This is likely to
underestimate the true field strength by a factor of $\sim2$, due to
biases induced by correlations between electron density and magnetic
field fluctuations in interstellar gas (\cite{bssw03}).

\subsection{The Random Magnetic Field}

In addition to the coherent field, Figure~\ref{fig_rms} suggests the
presence of significant fluctuations in RM from point to point --- the
standard deviation in RM over the LMC is 81~rad~m$^{-2}$.  On large
scales, a structure function analysis reveals a clear outer scale to
RM fluctuations at a scale of $\sim100$~pc. This may represent the
characteristic scale for the shells and bubbles of ionised gas which
dominate the LMC's morphology as seen in H$\alpha$ (e.g., \cite{mea80}).
We will show in a forthcoming paper that these measurements imply a
ratio of random to ordered magnetic field strengths on large scales of
$\approx3.6$, so that the random field dominates the ordered field.

\begin{wrapfigure}{l}{0.55\textwidth}
\vspace{-4mm}
\centerline{\psfig{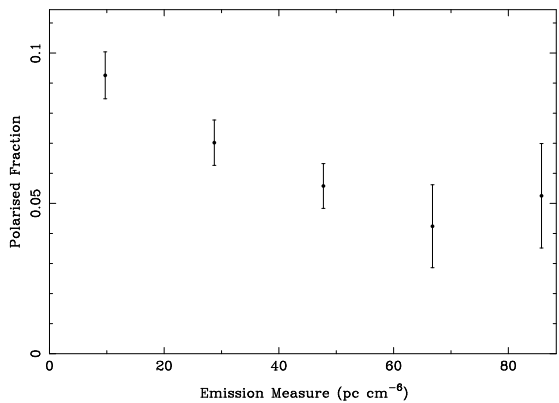}}
\vspace{-2mm}
\caption{Average polarised fraction vs. emission measure for polarised
sources behind the LMC.}
\label{fig_em}
\vspace{-2mm}
\end{wrapfigure}

Fluctuations in RM on much smaller scales are implied by the fact
that polarised sources seem to avoid bright H$\alpha$ regions
in Figure~\ref{fig_rms}.  At a quantitative level, the brightest
foreground H$\alpha$ intensity against which we obtain an RM is
$\approx75$~rayleighs, even though 7\% of all H$\alpha$ pixels within
the boundaries of the LMC are brighter than this.  About 90 of our
RMs lie directly against the LMC. If these sources are randomly
distributed, the probability of having none of them lie behind more
than 75~rayleighs of H$\alpha$ sources is $(1-0.07)^{90} \approx
1.5\times10^{-3}$. Another way of looking at this effect is to plot
polarised fraction vs. foreground emission measure (EM) for sources behind
the LMC. As shown in Figure~\ref{fig_em}, a clear trend is apparent,
such that sources behind high EMs seem to be increasingly depolarised.

For these observations this effect can only be due to beam depolarisation,
implying large fluctuations in RM on scales smaller than the angular
sizes of our sources. All our sources are unresolved, so we must
appeal to source statistics (e.g., \cite{wfpl93}). From such studies
we estimate a median angular size for our sample of $\approx6''$, or
$\approx1.5$~pc at the distance of the LMC. Significant RM fluctuations on
scales $\ll1.5$~pc are implied. In a forthcoming paper, we show that the
observed depolarisation can be explained by a random field of strength
5~$\mu$G on scales of $\sim0.5$~pc. These large field fluctuations do
not seem to be part of the standard turbulent cascade from large scales,
but rather seem to trace a separate source of fluctuations on sub-parsec
scales. Similar small-scale turbulence is seen in RM measurements in our
own Galaxy, and possibly delineates the effects of individual stars and
\HII\ regions on interstellar gas (e.g., \cite{hgm+04}).

\section{Future RM Experiments with the SKA}

The work presented above is a preview of the approach to magnetic field
studies that will be routine with the Square Kilometre Array (SKA),
a next-generation radio telescope currently being developed by the
international community.

%\begin{wrapfigure}{r}{0.57\textwidth}
\begin{figure}
%\centerline{\psfig{file=fig_gaensler4.eps,width=0.54\textwidth}}
\centerline{\psfig{file=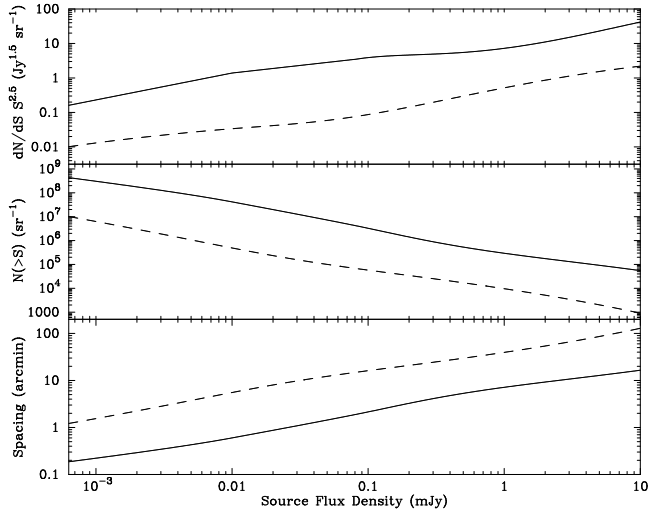,width=0.6\textwidth}}
\caption{Distribution of extragalactic source counts in both total
intensity (solid line) and in linear polarisation (dashed line) at an
observing frequency of 1.4~GHz (\cite{bg04}). The upper, middle and lower
panels show the predicted distribution of differential source counts,
integral source counts, and mean spacing between sources, respectively.
Surveys with current telescopes have a detection threshold of $\sim1$~mJy,
corresponding to a density of RM measurements of
$\sim1$~sources~deg$^{-2}$. An all-sky SKA survey will detect
sources down to $\sim1$~$\mu$Jy, corresponding to 
$\sim1500$~RMs deg$^{-2}$.}
\label{fig_grid}
\vspace{-2mm}
%\end{wrapfigure}
\end{figure}

``The Origin and Evolution of Cosmic Magnetism'' has been named
one of five key science projects for the SKA (Gaensler
\etal\ 2004\nocite{gbf04}).  It is
correspondingly highly likely that early in its operations, the SKA
will undertake an all-sky continuum survey, providing images of linear
polarisation at 1.4~GHz and at arcsecond resolution down to a sensitivity of
$\approx0.1$~$\mu$Jy. Among many exciting applications, this survey will
yield RMs from $\sim2\times10^7$ polarised extragalactic radio sources
(\cite{bg04}). From our prediction of the distribution of source counts
(see Fig.~\ref{fig_grid}), we expect that RMs will be distributed on
the sky at a mean spacing of only $\sim90''$ between sources!

The resulting ``RM grid'' will be a fantastic resource for probing
magnetic fields in all manner of foreground sources. For the LMC we can
expect in excess of $10^5$ background RMs, while we will obtain thousands
of RMs for many other nearby galaxies and clusters, in dramatic contrast
to the handful of RMs available for such sources now (e.g., Han
\etal\ 1998\nocite{hbb98};
\cite{gtd+01}).

On larger scales, this RM grid can be used to map the magnetic field of
the intergalactic medium (IGM) and of the overall Universe.  Magnetic
fields from the IGM, although yet to be detected, are likely to act
as the seed fields for galaxies and clusters, and may trace or even
regulate the formation of structure on large scales. Presently there
are dozens of models for the origin and distribution of IGM magnetic
fields, each model potentially making a different prediction for the
power spectrum of fluctuations in this large-scale field (for recent
reviews see \cite{gr01}; \cite{wid02}; \cite{gio04}).

Faraday rotation can potentially detect and map the magnetic field of
the IGM, but at the moment only upper limits have been obtained, at the
level of $10^{-8}$ to $10^{-9}$~G (\cite{kro94}; \cite{bbo99}).  Since
comoving electron densities and magnetic fields both increase 
in magnitude as a function of redshift, a survey which can accumulate
a large sample of RMs from high-redshift sources can potentially detect
the RM signature of the IGM (\cite{kol98}).  The SKA, in conjunction with
optical data from experiments such as LSST or SkyMapper, will have the
necessary sensitivity to make these measurements, ultimately yielding
the magnetic power spectrum of the observable Universe.

\section{Conclusions}

Our main results are as follows:

\begin{itemize}
\item The LMC has a coherent magnetic field with a spiral geometry, and
of strength $\ga1$~$\mu$G. 

\item On large ($\ga100$~pc) scales, there is also a significant random
component to the magnetic field, which dominates the ordered component
by a factor of $\sim3$.

\item On small ($\ll1$~pc) scales, there are strong spatial fluctuations
in rotation measure.

\end{itemize}

We next plan to interpret the diffuse polarised emission from the LMC
seen in Figure~\ref{fig_diffuse}(b). We also have recently obtained
corresponding ATCA data on the SMC and on part of the Magellanic Bridge,
to which we can apply a similar analysis to that presented here.

This work clearly demonstrates that background RMs, when obtained
at sufficiently high density on the sky, can be an exciting and
insightful way of mapping magnetic fields. With the advent of the SKA,
this will likely became a standard technique for studying magnetism at
all distances. \\

\small \noindent
We thank the organisers for running a particularly stimulating and
enjoyable conference.  We are grateful to Sungeun Kim for carrying
out the original ATCA observations which made this project possible.
The Southern H-Alpha Sky Survey Atlas (SHASSA) is supported by the
National Science Foundation.  The Australia Telescope is funded by the
Commonwealth of Australia for operation as a National Facility managed
by CSIRO.  B.M.G. acknowledges the support of the National Science
Foundation through grant AST-0307358, and of the University of Sydney
through the Denison Fund.

\normalsize

\bibliographystyle{apj1}
\bibliography{journals,modrefs,psrrefs}

\end{document}